\documentclass[twocolumn,prl]{revtex4}

\usepackage{graphicx}% Include figure files
\usepackage{dcolumn}% Align table columns on decimal point
\usepackage{bm}% bold math

\begin{document}
\title{Josephson effect as a measure of quantum fluctuations 
in the cuprates}

\author{R.~Hlubina, M.~Grajcar, and J.~Mr\'{a}z}

\affiliation{Department of Solid State Physics, Comenius University,
Mlynsk\'{a} Dolina F2, 842 48 Bratislava, Slovakia}

\begin{abstract}
Quantitative analysis of the Josephson effect is shown to provide
direct information about phase fluctuations in the superconducting
banks. Applying the analysis to the cuprates, substantial quantum
fluctuations between $d$-wave and $s$-wave pairing are found at low
temperatures.  A phenomenological model of such fluctuations is
introduced and solved in a self-consistent single-site
approximation. The cuprate parameters are argued to lie close to an
exotic superconductor phase with condensation of electron quadruplets.
\end{abstract}
\pacs{PACS}
\maketitle

More than a decade ago, it has been suggested that the Josephson
effect can be used as a direct probe of the angular dependence of the
order parameter phase in the cuprates \cite{Sigrist92}. Since then,
several types of phase sensitive experiments have been performed, and
an overwhelming majority of them confirmes the $d$-wave symmetry of
pairing in the cuprates (for a review, see \cite{Tsuei00}).

In absence of time reversal symmetry breaking, the superconducting
current $I$ as a function of the phase difference $\phi$ across the
Josephson junction reads $I(\phi)=\sum_n I_n \sin n\phi$
\cite{Likharev79}, and in the usually applicable tunnel limit only the
first harmonics $I_1$ needs to be considered.  In typical phase
sensitive experiments \cite{Tsuei00}, one tests only the dependence of
the sign of $I_1$ on the junction geometry.  In a recent series of
experiments the function $I(\phi)$, which contains a more complete
information about the equilibrium superconducting properties of a
Josephson junction, has been measured for junctions involving the
cuprates (for a review, see \cite{Hlubina02b}).

The present work is motivated by the special features of $I(\phi)$
found in the cuprates, in particular by the small first harmonics and
the large second harmonics of $I(\phi)$.  The outline of this Letter
is as follows.  First we show that a quantitative analysis of the
current-phase relation $I(\phi)$ provides information about the phase
fluctuations in the superconducting banks
\cite{Hlubina02a,Hlubina02b}. Further we introduce a phenomenological
model of phase fluctuations in the cuprates and solve it within the
simplest nontrivial self-consistent approximation.  Finally we discuss
the microscopic justification of the phenomenological model.

Let us start by considering a Josephson junction described, in absence
of phase fluctuations in the banks, by the harmonics $I_n$.  Now, let
us switch on phase fluctuations and denote the fluctuating part of the
phase in the bank $i=1,2$ as $\varphi_i$. In the tunnel limit the
phase fluctuations in different banks can be considered independent of
each other. Thus at time scales longer than the typical fluctuation
time and/or for junctions larger than the fluctuation correlation
length, $I(\phi)$ renormalizes to
$$
\tilde{I}(\phi)= 
\sum_n I_n\langle\sin n(\phi+\varphi_1-\varphi_2)\rangle=
\sum_n \tilde{I_n}\sin n\phi,
$$ 
where the renormalized $n$-th harmonic reads
$\tilde{I_n}/I_n=\langle\cos n\varphi_1\rangle \langle\cos
n\varphi_2\rangle$.  In grain boundary junctions with a $45^\circ$
misorientation between the cuprates, the first two harmonics are
experimentally accessible.  A detailed analysis of the data
\cite{Hlubina02a,Hlubina02b} suggests that
\begin{eqnarray}
\alpha=\langle\cos\varphi\rangle\approx 0.3,
\hspace{1cm}
\beta=\langle\cos 2\varphi\rangle\approx 1.
\label{eq:expt}
\end{eqnarray}
We would like to point out that the value of $\alpha$ was determined
from the Josephson product $I_cR_N$ of a large set of cuprate/cuprate
and of cuprate/low-$T_c$ junctions (for reviews, see
\cite{Hilgenkamp02,Tsuei00}) which are renormalized by $\alpha^2$ and
$\alpha$, respectively.  The values of $\alpha$ determined using both
types of junctions are in semiquantitative agreement with each other
and as such appear to be quite reliable \cite{Hlubina02a,Hlubina02b}.
On the other hand, the value of $\beta$ is subject to a larger
uncertainty, being based on a much smaller set of experimental data
\cite{Hlubina02b}. However, the (anomalous) inequality $\beta>\alpha$
seems to be quite reliable.

\begin{figure}
\centerline{\includegraphics[width=5.0cm,angle=0]{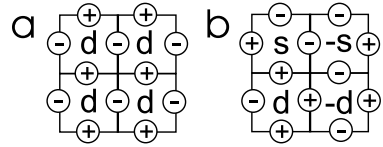}}
\caption{\label{fig:snapshot} (a) Average phase field configuration
$\langle\varphi_i\rangle$ (circles at the dual lattice sites) in a
$d$-wave superconductor.  Note the two sublattice structure of the
dual lattice.  (b) A typical snapshot of the fluctuating phase field
$\varphi_i=\langle\varphi_i\rangle+\delta_i$.  A lattice site $R$ (or
the corresponding lattice plaquette) is said to support local $d$,
$-d$, $s$, and $-s$ pairing, if phases at the bonds $\langle
R,R+x\rangle$ and $\langle R,R+y\rangle$ are $(+,-)$, $(-,+)$,
$(+,+)$, and $(-,-)$, respectively.}
\end{figure}

Now we proceed by introducing a minimal model of phase fluctuations
consistent with Eq.~(\ref{eq:expt}). Our model is motivated by the
strong-coupling RVB picture of the cuprates which views the cuprates
as a liquid of predominantly local singlets \cite{Anderson87}. While
the pairs are formed at a high energy scale $\sim J$, they acquire
phase coherence only at a much lower scale. Therefore we assign a
fluctuating phase field $\varphi_i$ to each bond $i$ of the square Cu
lattice. Note that $\varphi_i$ is the phase of a Cooper pair, i.e.
of a charge $2e$ object.

Let us first consider a toy model in which the fluctuating part
$\delta_i$ of the phase at each link is either 0 or $\pm\pi$, with
probabilities $P_1=(1+\alpha)/2\approx 0.65$ and
$P_2=(1-\alpha)/2\approx 0.35$, respectively. One finds readily that
Eq.~(\ref{eq:expt}) is satisfied in such a model.  On the other hand,
Fig.~\ref{fig:snapshot} shows a snapshot of a typical $\varphi_i$
configuration for such fluctuations.  Note that the probabilities that
a lattice point supports local $d$, $-d$, $s$, and $-s$ pairing
symmetry are $P_1^2\approx 0.42$, $P_2^2\approx 0.12$, $P_1P_2\approx
0.23$, and $P_1P_2\approx 0.23$, respectively.  In other words, the
numerical values in Eq.~(\ref{eq:expt}) suggest the presence of strong
quantum fluctuations from $d$-wave to $s$-wave pairing
\cite{Hlubina02a}.

The simplest model of a superconducting CuO$_2$ plane which allows for
phase fluctuations, takes into account the compact nature of the phase
field, and assumes dominant $d$-wave and subdominant $s$-wave pairing
(with pairing strengths $V+W$ and $V-W$, respectively), reads
\begin{eqnarray}
H=-m^{-1}\sum_i{\partial^2/\partial\varphi_i^2}
+\sum_{\langle i,j\rangle} E_J(\varphi_i,\varphi_j).
\label{eq:model}
\end{eqnarray}
The first term in Eq.~(\ref{eq:model}) which allows for quantum
fluctuations of phase can be thought of as a local charging energy at
site $i$.  Note that from now on, instead of bonds we talk about the
sites of a (dual) lattice.  The second term where $\langle i,j\rangle$
denotes a pair of nearest neighbor sites can be viewed as a Josephson
coupling between the sites $i$ and $j$, $E_J(\varphi_i,\varphi_j)=
-V\cos(2\varphi_i-2\varphi_j)+W\cos(\varphi_i-\varphi_j)$.  Note that
the sign of the $W$ term is different from the standard convention,
because of the $d$-wave symmetry of pairing in the cuprates.

Let us point out that our model Eq.~(\ref{eq:model}) includes the
quantum XY model \cite{Chakravarty88} as a special limit $V=0$; we
will show that the most spectacular effects occur at $V\gg W$. Note
also that we haven't considered dissipative effects in
Eq.~(\ref{eq:model}).  Since the cuprates are $d$-wave superconductors
with low-lying fermionic excitations \cite{Lee97}, this is a
nontrivial assumption which needs to be studied in future work.

In what follows we study the model Eq.~(\ref{eq:model}) by
constructing a variational solution of the form
\begin{equation}
\psi(\varphi_1,\varphi_2,\ldots)=\Pi_i\psi_i(\varphi_i),
\label{eq:wavefun}
\end{equation}
where the product runs over all lattice sites and the local functions
$\psi_i$ are $2\pi$ periodic.  The Schr\"odinger equation for the
lattice point $i$ reads
\begin{equation}
[-m^{-1}{\partial^2/\partial\varphi^2}+U_i(\varphi)]
\psi_i(\varphi)=\varepsilon_i\psi_i(\varphi),
\label{eq:schrodinger}
\end{equation}
with a self-consistent potential
$$
U_i(\varphi)=\sum_\tau
\left[-V\langle\cos 2\varphi\rangle_{i+\tau}\cos 2\varphi
+W\langle\cos\varphi\rangle_{i+\tau}\cos\varphi\right],
$$ 
where the sum runs over the four nearest-neighbor sites $i+\tau$ of
the studied site and $\langle f(\varphi)\rangle_i=\int d\varphi
f(\varphi) |\psi_i(\varphi)|^2$.  Note that the single-site
approximation Eq.~(\ref{eq:wavefun}) is similar in spirit to the
self-consistent harmonic approximation \cite{Chakravarty88}, but goes
beyond it by treating the local problem exactly, which is instrumental
in the present context.

Now we specialize to the case of a two-sublattice solution with
$\psi_i(\varphi)=\psi_A(\varphi)$ and
$\psi_i(\varphi)=\psi_B(\varphi)$ for $i$ in sublattice $A$ and $B$,
respectively.  Note that this corresponds to a translationally
invariant solution on the original lattice.  Assuming furthermore a
$d$-wave solution, $\psi_B(\varphi)=\psi_A(\varphi+\pi)$, we obtain
$\langle\cos n\varphi\rangle_B=(-1)^n\langle\cos n\varphi\rangle_A$.
The self-consistent Hamiltonian for sublattice $A$ therefore reads
\begin{equation}
H=-{\partial^2/\partial\varphi^2}-a\cos\varphi-b\cos 2\varphi,
\label{eq:self1}
\end{equation}
where we have measured the energy in units of $m^{-1}$.  Introducing
dimensionless interaction parameters $v=mV$ and $w=mW$, the
dimensionless self-consistent potentials read
\begin{eqnarray}
a=4w\alpha,
\hspace{2cm}
b=4v\beta, 
\label{eq:self2}
\end{eqnarray}
where $\alpha$ and $\beta$ are defined in Eq.~(\ref{eq:expt}) in which
the mean values $\langle\ldots\rangle$ are to be calculated with
respect to the solution of Eq.~(\ref{eq:self1}).  The total energy per
lattice site ${\bar\varepsilon}$ differs from the lowest eigenvalue
$\varepsilon$ of the Hamiltonian Eq.~(\ref{eq:self1}):
\begin{equation}
{\bar\varepsilon}=\varepsilon+(a\alpha+b\beta)/2.
\label{eq:energy}
\end{equation}

\begin{figure}
\centerline{\includegraphics[width=6.0cm,angle=0]{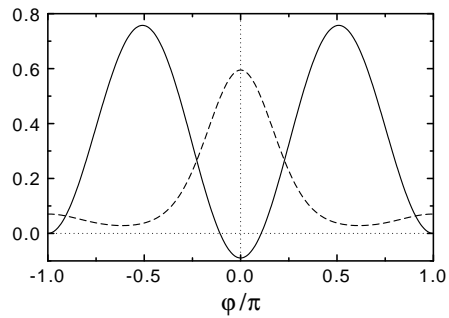}}
\caption{\label{fig:wavefun} Solid line: Potential energy entering
Eq.~\ref{eq:self1} (in arbitrary units). Result of a self-consistent
calculation for $v=1.1$ and $w=0.09$, leading to $a\approx 0.22$ and
$b\approx 2.0$. Dashed line: Square of the ground-state wavefunction
$|\psi(\varphi)|^2$ for the same parameters.  }
\end{figure}

In Fig.~\ref{fig:wavefun} we plot the potential energy entering
Eq.~(\ref{eq:self1}) in the limit $b\gg a$ relevant to the
cuprates. The classical solutions at $\varphi=0$ and $\varphi=\pm\pi$
are nearly degenerate, their splitting being $2a$.  The ground-state
wavefunction $\psi(\varphi)$ is localized predominantly in the
vicinity of $\varphi=0$, but tunneling across the finite barrier leads
to a second peak of $\psi(\varphi)$ in the vicinity of
$\varphi=\pm\pi$.  In complete analogy with the toy model
Fig.~\ref{fig:snapshot}, this tunneling leads to quantum fluctuations
between $d$-wave and $s$-wave pairing, which are thus seen to be
essentially local.

\begin{figure}
\centerline{\includegraphics[width=6.0cm,angle=0]{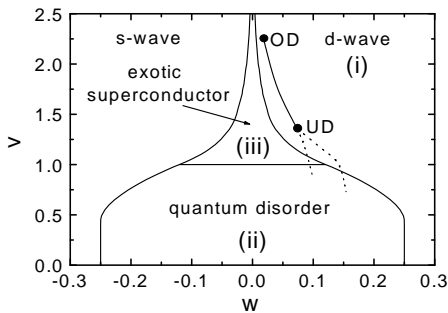}}
\caption{\label{fig:phase_diag} Phase diagram of the model
Eq.~(\ref{eq:model}). We hypothesize that superconducting cuprates
track the line between the points OD and UD, corresponding to
overdoped and underdoped materials. The dotted lines show possible
trajectories under further reducing the hole density.}
\end{figure}

Self-consistent solution to
Eqs.(\ref{eq:self1},\ref{eq:self2},\ref{eq:expt}) shows that three
phases exist in the $w>0$ part of the phase diagram of the model
Eq.~(\ref{eq:model}), see Fig.~\ref{fig:phase_diag}: (i) a
conventional $d$-wave superconducting phase with $\alpha>0$ and
$\beta>0$ realized for sufficiently large $v,w$; (ii) a quantum
disordered phase with $\alpha=\beta=0$ at small $v,w$; and (iii) an
exotic superconductor phase with $\alpha=0$ and $\beta>0$ appearing at
large $v$ and small $w$.

Extending the phase diagram for $w>0$ to negative values of $w$, one
verifies easily that the role of $d$-wave and $s$-wave pairing is
interchanged and therefore the phase diagram for $w<0$ is a mirror
image of that for $w>0$. This implies that within the quantum model
Eq.~(\ref{eq:model}), upon changing the sign of $W$ the
superconducting state switches from $d$-wave to $s$-wave symmetry by
crossing one of the two nontrivial phases in
Fig.~\ref{fig:phase_diag}. Which of the two possible intermediate
states is realized depends on the strength of quantum fluctuations. In
the limit of strong fluctuations (small $v$), an intermediate quantum
disordered state is realized [phase (ii)], which is presumably
insulating.  If the strength of quantum fluctuations decreases, the
exotic phase (iii) separates the two conventional pairing states. In
the classical limit $m\rightarrow\infty$ the width of phase (iii)
vanishes. In what follows we argue that although
$\langle\cos\varphi\rangle=0$ in phase (iii), it does represent a
superconductor, albeit an exotic one, since it can be thought of as a
condensate of pairs of Cooper pairs, or quadruplets of electrons
\cite{note}. Note that phase (iii) bears some similarity to the
fractionalized superconductor discussed in \cite{Senthil00}.

\begin{figure}
\centerline{\includegraphics[width=6.0cm,angle=0]{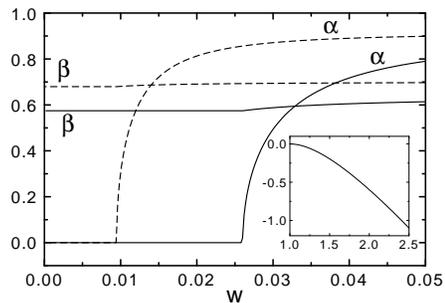}}
\caption{\label{fig:cut} Order parameters $\alpha$ and $\beta$ along
two cuts of the phase diagram Fig.~\ref{fig:phase_diag}: $v=1.5$
(solid lines) and $v=2$ (dashed lines). The inset shows the total
energy per lattice site $\bar{\varepsilon}$ as a function of $v$ along
the line $w=0$.}
\end{figure}

Let us show now that phase (iii) is described by a macroscopic
Ginzburg-Landau type wavefunction $\Psi_{GL}(x,y)$ with a nonvanishing
phase rigidity. To this end consider the many body wavefunction
Eq.~(\ref{eq:wavefun}) with the following choice of local
wavefunctions,
\begin{eqnarray}
\psi_{i+\hat{x}}(\varphi)&=&\psi_{i}(\varphi+\pi+\chi_1), 
\nonumber\\
\psi_{i+\hat{y}}(\varphi)&=&\psi_{i}(\varphi+\pi+\chi_2),
\label{eq:psi_modulated}
\end{eqnarray}
where $\hat{x}$ and $\hat{y}$ are unit vectors of the lattice in the
$x$ and $y$ directions, respectively. Eq.~(\ref{eq:psi_modulated})
corresponds to a macroscopic wavefunction with an internal $d$-wave
symmetry and a spatially modulated phase
$\Psi_{GL}(x,y)\propto\exp{-i(\chi_1x+\chi_2y)/d_0}$, where $d_0$ is
the lattice constant of the dual lattice. The Schr\"odinger equation
for site $i$ can still be written in the form of
Eq.~(\ref{eq:schrodinger}) with a selfconsistent potential satisfying
$U_{i+\hat{x}}(\varphi)=U_{i}(\varphi+\pi+\chi_1)$ and
$U_{i+\hat{y}}(\varphi)=U_{i}(\varphi+\pi+\chi_2)$.  The periodicity
of the local wavefunctions together with Eq.~(\ref{eq:psi_modulated})
guarantees that all local energies $\varepsilon_i$ are equal.  For a
general lattice point $i$ the potential $U_i(\varphi)$ contains terms
proportional to both $\cos n\varphi$ and $\sin n\varphi$ with
$n=1,2$. However, we assume that there exists one special lattice
point for which the terms propoportional to $\sin n\varphi$
vanish. For this lattice point the Hamiltonian can be written in the
form of Eq.~(\ref{eq:self1}).  The set of self-consistent equations is
closed by Eq.~(\ref{eq:self2}), in which the effective interaction
strengths $w_{\rm eff}= w(\cos\chi_1+\cos\chi_2)/2$ and $v_{\rm
eff}=v(\cos 2\chi_1+\cos 2\chi_2)/2$ have to be used instead of $w$
and $v$, respectively.  Since the total energy per lattice site is
still given by Eq.~(\ref{eq:energy}), the energy difference between a
modulated and a uniform macroscopic state can be written in a
Ginzburg-Landau type form
\begin{equation}
\delta\bar{\varepsilon}=
\bar{\varepsilon}(w_{\rm eff},v_{\rm eff})-\bar{\varepsilon}(w,v)
\approx -{\partial\bar{\varepsilon}\over\partial w}\delta w
-{\partial\bar{\varepsilon}\over\partial v}\delta v,
\label{eq:GL}
\end{equation}
where $\delta w=w-w_{\rm eff}\approx w(\chi_1^2+\chi_2^2)/4$ and
$\delta v=v-v_{\rm eff}\approx v(\chi_1^2+\chi_2^2)$ and the
approximate equations are valid for slowly varying macroscopic
wavefunctions, $\chi_1,\chi_2\ll 1$. In the conventional
superconductor phase [phase (i)] we expect in a generic point of the
phase diagram $\partial\bar{\varepsilon}/\partial w<0$ and
$\partial\bar{\varepsilon}/\partial v<0$. From Eq.~(\ref{eq:GL}) it
then follows that in the long-wavelength limit
$\delta\bar{\varepsilon} \propto (\chi_1^2+\chi_2^2)$, as it should be
in a superconductor with a finite phase stiffness. On the other hand,
in phase (iii) we have $\partial\bar{\varepsilon}/\partial w=0$, but
the inset to Fig.~\ref{fig:cut} shows explicitly that
$\partial\bar{\varepsilon}/\partial v<0$, which is enough to guarantee
a finite phase stiffness in this case as well. Thus we have shown that
the phase (iii) is an exotic superconductor.

Returning to the physics of cuprates, Eq.(\ref{eq:expt}) implies that
if we insist on their description in terms of the model
Eq.~(\ref{eq:model}), then its parameters should be chosen inside the
phase (i), but close to phase (iii). Fig.~\ref{fig:cut} shows
explicitly that in this case it is possible to have $\alpha\approx
0.3<\beta$.  Within our simplistic model it is difficult to have
$\beta\approx 1$ at the same time, because this would require small
quantum fluctuations (large $v$).  However, for $v\gg 1$ the
difference $W$ between the $d$-wave and $s$-wave pairing strengths
needs to be implausibly small in order to stabilize the exotic
superconductor phase.

\begin{figure}
\centerline{\includegraphics[width=6.0cm,angle=0]{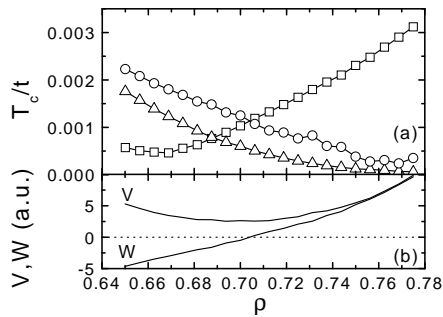}}
\caption{\label{fig:Tc} (a) Superconducting transition temperature of
the $t$-$t^\prime$ Hubbard model in the $d$-wave ($T_{cd}$, squares)
and $s$-wave ($T_{cs}$, circles) symmetry sectors for
$t^\prime/t=0.45$ and $U/t=6$ as a function of electron density. Also
shown is the transition temperature $T_{cxy}$ in the leading
subdominant symmetry sector ($d_{xy}$-wave, triangles).  Calculated
following \cite{Mraz02} on a lattice 1024$\times$1024. (b) Coupling
constants $V$ and $W$ (in arbitrary units) estimated from $T_{cd}$ and
$T_{cs}$.}
\end{figure}

Next we address the question whether it is reasonable to assume that,
within a realistic model of the cuprates, $d$-wave and $s$-wave
pairing are competitive in energy.  Lead by the hypothesis that the
symmetry of pairing is correctly determined already at weak coupling
\cite{Hlubina99}, we have studied the superconducting phase diagram of
the $t$-$t^\prime$ Hubbard model for $t^\prime/t=0.45$ and a moderate
interaction strength $U/t=6$ as a function of the electron density
$\rho$ in the vicinity of $\rho=0.7$. This choice of parameters is
close to the canonical values for the cuprates and it is inspired by
the results of \cite{Hlubina99}, where a $d$-wave/$s$-wave pairing
transition was found in this region.

In Fig.~\ref{fig:Tc} we plot the superconducting transition
temperatures determined by the variational method for a Kohn-Luttinger
superconductor \cite{Mraz02} in the two dominant pairing symmetries
($d$-wave and $s$-wave). Note that for extremely overdoped cuprates
with $\rho<\rho_c\approx 0.7$, $s$-wave pairing is the leading
instability.  It is also shown that close to $\rho_c$, $T_c$ in the
leading subdominant pairing symmetry sector ($d_{xy}$-wave) is much
smaller than $T_{cd}$, $T_{cs}$.

According to Eq.~(\ref{eq:model}), the condensation energy per site in
absence of phase fluctuations is $2(V+W)$ and $2(V-W)$ in the $d$-wave
and $s$-wave pairing sectors, respectively. At weak coupling, these
condensation energies are proportional to $T_{cd}^2$ and
$T_{cs}^2$. By comparing these expressions, we obtain a rough estimate
of the coupling constants $V$ and $W$.  The results are shown in
Fig.~\ref{fig:Tc}.  Note that with hole doping $x=1-\rho$ decreasing
from $x_c\approx 0.3$ towards experimentally accessible values, both
$W$ and $V$ and their ratio $W/V$ increase.  On the other hand, also
the fluctuations are expected to grow with underdoping,
i.e. $m\rightarrow 0$ as $x\rightarrow 0$
\cite{Anderson87,Uemura89}. This brings us to the question about the
doping dependence of $v=mV$ and $w=mW$. We hypothesize that with
reduced doping, the effect of $m\rightarrow 0$ dominates and the
cuprates track the line in Fig.~\ref{fig:phase_diag} between the
points OD and UD.  Under further reducing the hole density, the
insulating quantum disordered phase should be reached. This can happen
either directly or via crossing the exotic superconductor phase, as
indicated by the dashed lines in Fig.~\ref{fig:phase_diag}.

It should be remarked that the question about how $I(\phi)$ depends on
the doping level is very difficult to address experimentally.  An
interesting possibility might be to study artificially doped grain
boundary junctions \cite{Hammerl00}.

In conclusion, we have found that the two seemingly independent
experimental observations in the cuprates, namely the large first and
the small second harmonics of the Josephson current, can be explained
by a single assumption of substantial quantum fluctuations between
$d$-wave and $s$-wave pairing at low temperatures.  We have introduced
a phenomenological model of such fluctuations and solved it in a
self-consistent single-site approximation. We have argued that the
cuprate parameters lie close to an exotic superconductor phase with
condensation of electron quadruplets.  Finally, $d$-wave and $s$-wave
pairing were shown to be competitive within the Hubbard model for
parameters close to the canonical values for the cuprates.

R.~H. thanks E.~Tosatti and T.~V.~Ramakrishnan for useful comments.
This work was supported by the Slovak Scientific Grant Agency under
Grant No.~VEGA-1/9177/02 and by the Slovak Science and Technology
Assistance Agency under Grant No.~APVT-51-021602.


\begin{thebibliography}{99}
\bibitem{note} For nonlocal pairs
$b^\dagger_{ij}=(c^\dagger_{i\uparrow}c^\dagger_{j\downarrow}-
c^\dagger_{i\downarrow}c^\dagger_{j\uparrow})/\sqrt{2}$, quadruplet
condensation is possible since
$(b^\dagger_{ij})^2=-c^\dagger_{i\uparrow}c^\dagger_{i\downarrow}
c^\dagger_{j\uparrow}c^\dagger_{j\downarrow}$.
\end{thebibliography}
\end{document}